\RequirePackage{fix-cm}
\documentclass[smallextended]{svjour3}       
\smartqed  
\usepackage{graphicx}
\usepackage{amssymb}
\usepackage{amsmath}
\usepackage{soul}
\usepackage{enumerate}
\usepackage{color,xcolor}
\usepackage{threeparttable}
\usepackage{hyperref}
\usepackage[misc]{ifsym}
\usepackage{caption}
\usepackage{float}
\usepackage{subcaption}
\begin{document}
\title{A Meta Path Based Evaluation Method for Enterprise Credit Risk

}



\author{Marui Du \textsuperscript{1} 
 \and Yue Ma \textsuperscript{2}               
\and Zuoquan Zhang \textsuperscript{1}
}

\authorrunning{} 

\footnotetext[1]{School of Science, Beijing Jiaotong University, China}
\footnotetext[2]{Guanghua School of Management, Peking University, China}

\institute{Marui Du \at \email{17118446@bjtu.edu.cn} }

\date{Received: date / Accepted: date}
\maketitle

\begin{abstract}
Nowadays small and medium-sized enterprises have become an essential part of the national economy. 
With the increasing number of such enterprises, how to evaluate their credit risk becomes a hot issue.
Unlike big enterprises with massive data to analyze, it is hard to find enough information of small enterprises to assess their financial status.
Limited by the lack of primary data, how to inference small enterprises's credit risk from secondary data, like information of their upstream, downstream, parent and subsidiary enterprises, attracts big attention from industry and academy.
Targeting on accurately evaluating the credit risk of the small and medium-sized enterprise (SME), in this paper, we exploit the representative power of Information Network \cite{gupta2017heteclass} on various kinds of SME entities and SME relationships to solve the problem.
A novel feature named meta path feature proposed to measure the credit risk, which makes us able to evaluate the financial status of SMEs from various perspectives.
Experiments show that our method is effective to identify SME with credit risks.

\end{abstract}

\section{Introduction}
\label{intro}
Small and medium-sized enterprise (SME) is one of backbones in the national economy, whose development directly affects it. 
However, due to the incomplete management system and the lack of appropriate financial indicators, the credit risk assessment process is usually time-consuming, and the evaluation result is often of low accuracy. 
Therefore, in this paper, we are going to propose an appropriate method of credit risk assessment to target this problem.

Industry and academy always have a critical focus on how to measure enterprise credit risk.
Conventional approaches of assessment  mainly extract enterprise-related features, such as financial indicators, to predict enterprise solvency.
However, with the expansion of global market size in recent years, conventional approaches have lost their power of discrimination in the situations, where relations and interactions between SMEs are numerous and complicated. 
An SME's financial status can be easily affected by some actions from its related other SMEs.
For example, the contagion risk is caused by associated credit entities, which besets many SMEs with the risk of default even in good financial conditions.
Therefore, rather than single financial indicators, relations and interactions between SMEs should be paid more attention in studying SME credit risk.

To model the relations and interactions, various entities and their relationships can be considered in the information networks \cite{gupta2017heteclass}.
In the previous, most of researchers studied the above problem with a homogeneous information network \cite{sun2013mining} consisting only one single relation type and one entity type.
However in SME setting, the structure of homogeneous information network may be a bit simple to explain the relationships between SMEs.
To not lose important information, a heterogeneous information network \cite{2016A} with complicated graph structure is more suitable to study the interaction between SMEs.
In the heterogeneous information network, meta paths (MP) \cite{2016A} are taken as a fundamental data structure to capture semantical relationships between entities.
Through MP, complicated relationships between entities can be systematically and concisely defined.
The path provides a clear view of how entities interact mutually in the information network.
In this paper, to assess the status of SMEs, we exploit the power of meta path to study how influences among financial entities spread in the information network of SMEs. 

\begin{figure}[pth]
      \begin{center}
             \includegraphics[width=3.5in]{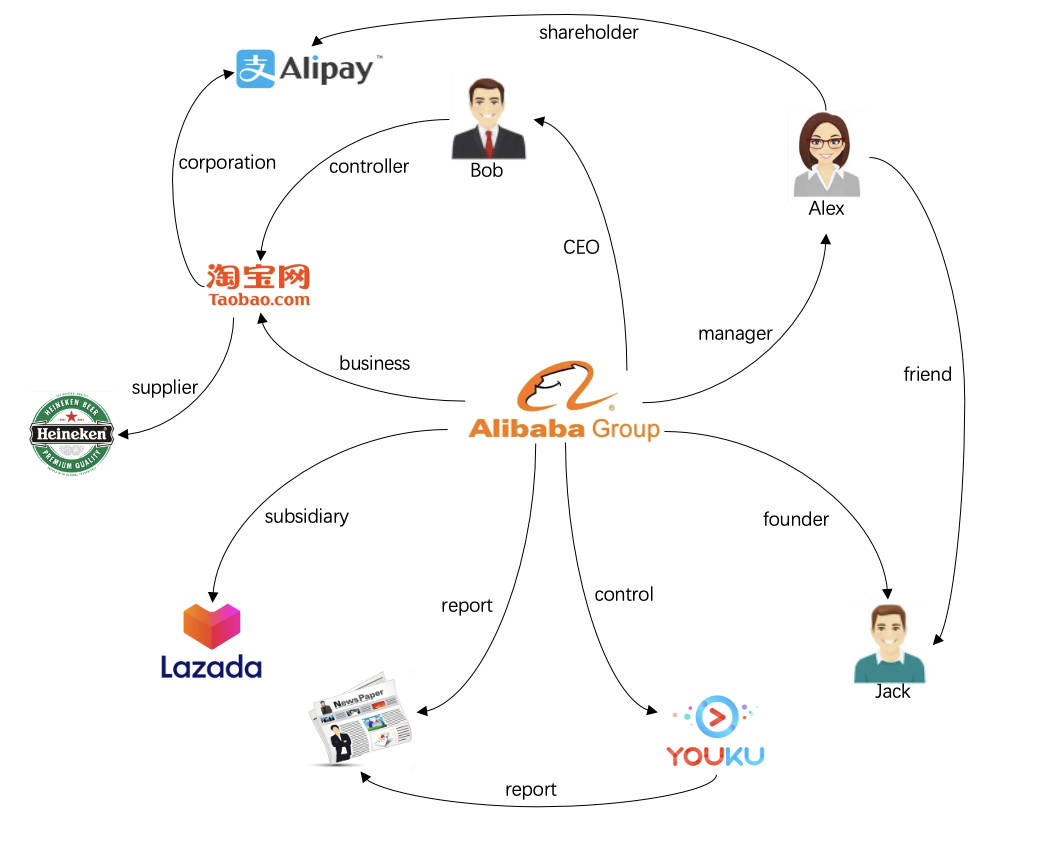}\newline
      \end{center}
             \caption{Alibaba heterogeneous information network example. There exist multiple types of nodes in the network, such as \emph{enterprise} (Alibaba, Lazada, YouKu, Heineken), \emph{person} (Bob, Alex, Jack), \emph{commodity} (Taobao, Alipay) and \emph{news} (newspaper).  Links between nodes represent relation connect entities, for example, Jack is the \emph{founder} of Alibaba, Heineken is the \emph{supplier} of Taobao, and the newspaper \emph{report} a piece of news of  Alibaba's \emph{control} company YouKu etc.}
             \label{fig:SMEs-HIN}
\end{figure}

In our method,  we first build a heterogeneous information network of SMEs to describe interactive relationships between different entities associated with SME.
Figure \ref{fig:SMEs-HIN} is a toy example of Alibaba heterogeneous information network, which demonstrates some possible connections of Alibaba and its related entities. 
For example, path ``$Alibaba\xrightarrow{subsidiary} Lazada$'' represents information that Lazada is a subsidiary of Alibaba; path ``$Alibaba \xrightarrow{CEO}{Bob}\xrightarrow{controller}{Taobao}$.'' represents information that Alibaba's CEO, Bob, is also Taobao's controller; and path ``$Alibaba\xrightarrow{control}{YouKu}\xrightarrow{report}{news}$.'' represents information that Alibaba's control enterprise, YouKu, is criticized by the newspaper.
It is easy to see that through information networks, the interrelated relations between entities can be easily obtained.
By building information network of SMEs, we can not only obtain the self-related information but also the interactive information associated with the target enterprise.

With the given information network of SMEs,  we propose a novel feature - meta path feature, to measure the impact through meta paths from one financial entity to another.
Unlike conventional financial indicators, the meta path feature can be defined and applied very flexibly.
The flexibility makes us able to evaluate the credit status of SMEs from various perspectives more comprehensively.
The proposed meta path feature can also explicitly show how much one entity can be affected by a specific logical path, which can provide an intuitive view for banks, lenders, and relevant experts to understand the credit risk faced by SMEs.
In this way, SME default can be effectively identified.

In the rest of this paper, section \ref{sec:related} introduces the SME credit risk evaluation method and the application of information networks. 
Section \ref{sec:model} builds a model of SMEs' heterogeneous information network and proposes the meta path feature. 
In section \ref{sec:impact}, by considering the ability of risk identification, three features are proposed based on meta path.
Section \ref{sec:experiments} presents the experiment on three real-world datasets and section \ref{sec:Conclusion} concludes the paper.

\section{Related work}
\label{sec:related}
The credit risk evaluation model of SMEs was first established by Edmister \cite{edmister1972empirical} in 1972, leading to the emergence of a large number of credit risk measurement index systems.
Most of the early credit evaluation models for SME at home and abroad follow the index system of credit evaluation model for large enterprises, that is, the extraction of some key financial indicators of enterprise financial statements. 
Among these key financial indicators, profitability indicators \cite{cultrera2016bankruptcy}, such as operating profit ratio and ratio of profits to cost, and solvency indicators \cite{tian2015variable}, such as current and quick ratio, are used the most. 
Besides, operational capacity indicators \cite{bauer2014hazard}, development capacity indicators \cite{sermpinis2018modelling}, and liquidity indicators \cite{sermpinis2018modelling} are added in many studies. 
Since financial indicators alone can not lineate the complete picture of an enterprise, non-financial indicators such as, managers background \cite{2010Evaluation}, working experience \cite{2013Feature}, and enterprise internal structure \cite{2010Predicting} \cite{moro2013loan} are added for evaluation. 
However, financial and non-financial indicators can not capture the contagion credit risk among financial entities since they are independent and do not consider the casual chain.
  
Recently, with the rapid improvement of computing capacity and the development of data mining technology, information network has gained much attention from researchers and makes excellent work in the field of clustering \cite{sun2009ranking} \cite{sun2012relation}, classification \cite{ji2010graph} \cite{wang2016text}, relation prediction \cite{sun2012will} \cite{popescul2003statistical} and recommendation \cite{shi2015semantic} \cite{ma2009learning}.
Researchers often use two kinds of information networks: the homogeneous information network and the heterogeneous information network.
The homogeneous information network builds with same type of objects and link relations.
For example, Jamali \cite{jamali2010matrix} builds a social network for user recommendation based on user ratings; 
Ma \cite{ma2008sorec} builds a friend relationship prediction network based on personal relations.
These homogeneous information networks ignore the relationship between different objects and relations, which causes the loss of important information.
The concept of heterogeneous information network was first proposed by Sun \cite{2016A} in 2009.
It combines more information and contains logical semantics of different object types and link types.
For example, Wang \cite{wang2018shine} proposes a Signed Heterogeneous Information Network Embedding to capture the sentiment links of online social information by considering users with sentiment and social relations; Hosseini \cite{hosseini2018heteromed} used the heterogeneous information network with high dimensional data and rich relationships for medical diagnosis.
The heterogeneous information network is usually used to capture complicated semantic and logical relationships among different entities.

In the field of SME credit risk evaluation, a large amount of data related to enterprises has been accumulated, such as upstream and downstream enterprise information and relevant news information. 
The heterogeneous relationships between different entities have also provided researchers with new ideas to find SME credit risk factors.
For example, Tsai \cite{tsai2017risk} pays attention to the impact of enterprise-related news information on the credit risk of SMEs; 
Yin \cite{yin2020evaluating} utilizes legal judgments to evaluate the credit risk of SMEs. 
Moro \cite{moro2013loan} takes the impact of SMEs and bank manager trust relationship on enterprise credit risk into consideration. 
Tobback \cite{2017Bankruptcy} pays attention to the inter-enterprise relationship data for feature selection to measure the credit risk of SMEs; 
Kou \cite{2020Bankruptcy} is focused on enterprise payment and transaction data information to measure the credit risk of SMEs. 
However, all of their works are built on homogeneous information networks, most of which do not consider the heterogeneous information.
Therefore, in this paper, we build a heterogeneous information network for SMEs to more effectively evaluate SME credit risks, which considers both the heterogeneous information of SMEs and the semantic information carried by different SME entities.

\section{Model of SME Credit Risk}\label{sec:model}
To evaluate SME credit risk, conventional methods adopted by experts usually make their judgments only based on the features directly affecting SME default, such as asset-liability ratio, current ratio, and turnover rate, but not on logical relationships between SMEs, such as parent and subsidiary situations, upstream and downstream situations, enterprise director or high-level manager related situations.
For example, when a parent company defaults, the solvency of its subsidiaries will also be affected.
If the influences exerted by the parent company are neglected, its subsidiary company's default conditions will be overestimated.
Therefore, apart from the features directly affecting default, the logical relationships between SMEs should also be considered in evaluating SMEs' status.
Paying attention to different connections between SMEs can improve both the reliability and the interpretability of the evaluation.
This section will give a model of SME credit risk with logical relationships adopted.

\subsection{SME Heterogeneous Information Network}\label{subsec:def}

A heterogeneous information network \cite{2016A} is a classical data structure used to model objects and relations in a directed graph.
This graph structure has shown its superiority in representing and storing knowledge about the natural world for many applications \cite{2013Modeling} \cite{2013Recommendation} \cite{2019Cash}.
Given different objects in information networks, logical connections can be effectively constructed, and semantic relationships can be easily captured.
Hence, we also build our model in a information network which is defined as follows:
\begin{definition}\label{def:graph}
With a schema $S=(\mathcal{A},\mathcal{R})$, an \emph{information network} defined as a directed graph $G=(\mathcal{V},\mathcal{E})$ with object type function $\tau:\mathcal{V}\rightarrow\mathcal{A}$ and relation type function $\phi:\mathcal{E}\rightarrow\mathcal{R}$, where object $v\in\mathcal{V}$ belongs to object type $\tau(v)\in\mathcal{A}$, link $e\in\mathcal{E}$ belongs to relation type $\phi(e)\in\mathcal{R}$.
\end{definition}

\begin{figure}[h]
      \begin{center}
            \includegraphics[width=3in]{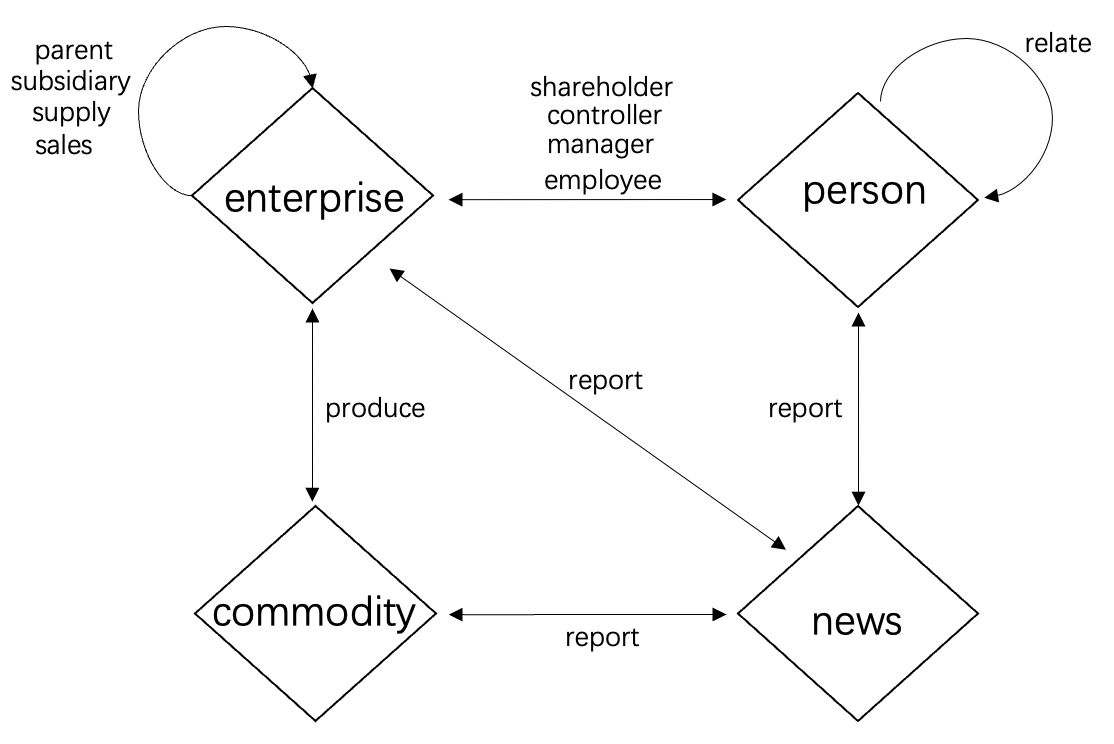}\newline
      \end{center}
             \caption{The SME network schema}
             \label{fig:schema}
\end{figure}

In this paper, our model is built as a heterogeneous information network of SMEs.
The SME schema is shown in Figure \ref{fig:schema}.

In our model, \emph{enterprise} ($\mathcal{A}_{e}$) , \emph{commodity} ($\mathcal{A}_{c}$), \emph{person} ($\mathcal{A}_{p}$) and \emph{news} ($\mathcal{A}_{n}$) are four fundamental object types in studying SME credit risk.
The studied relation types are summarized from public enterprise information and objective facts, such as the \emph{shareholder} relation between enterprise and person, the \emph{produce} relation between enterprise and commodity, and the \emph{report} relation between enterprise and news.
The types mentioned in this paper are listed in Table \ref{tab:den}. 
\begin{table}
\caption{Object type and relation type notations}
\label{tab:den}       
\centering
\begin{tabular}{cll}
  \hline\noalign{\smallskip}
     Notation & Discriptions \\
     \noalign{\smallskip}\hline\noalign{\smallskip}
                    $\mathcal{A}_{e}$ & the object type of \emph{enterprise}\\
                    $\mathcal{A}_{c}$ & the object type of \emph{commodity}\\
                    $\mathcal{A}_{p}$ & the object type of \emph{person}\\
                    $\mathcal{A}_{n}$ & the object type of \emph{news}\\
                    $\mathcal{R}_{parent}$ & the relation type of \emph{parent} between enterprises\\
                    $\mathcal{R}_{subsidiary}$ & the relation type of \emph{subsidiary} between enterprises\\
                    $\mathcal{R}_{supplyer}$ & the relation type of \emph{supply} between enterprises\\
                    $\mathcal{R}_{saler}$ &  the relation type of \emph{sales} between enterprises\\
                    $\mathcal{R}_{control}$ & the relation type of \emph{controller} between enterprise and person\\
                    $\mathcal{R}_{shareholder}$ & the relation type of \emph{shareholder} between enterprise and person\\
                    $\mathcal{R}_{manager}$ &  the relation type of \emph{manager} between enterprise and person\\
                    $\mathcal{R}_{employee}$ &  the relation type of \emph{employee} between enterprise and person\\
                    $\mathcal{R}_{produce}$ & the relation type of \emph{produce} between enterprise and commodity\\
                    $\mathcal{R}_{report}$ & the relation type of \emph{report} between enterprise and news\\
                    $\mathcal{R}_{relate}$ & the relation type of \emph{relate} between person\\
  \noalign{\smallskip}\hline
\end{tabular}
\end{table}

\begin{figure}[h]
      \begin{center}
             \includegraphics[width=3in]{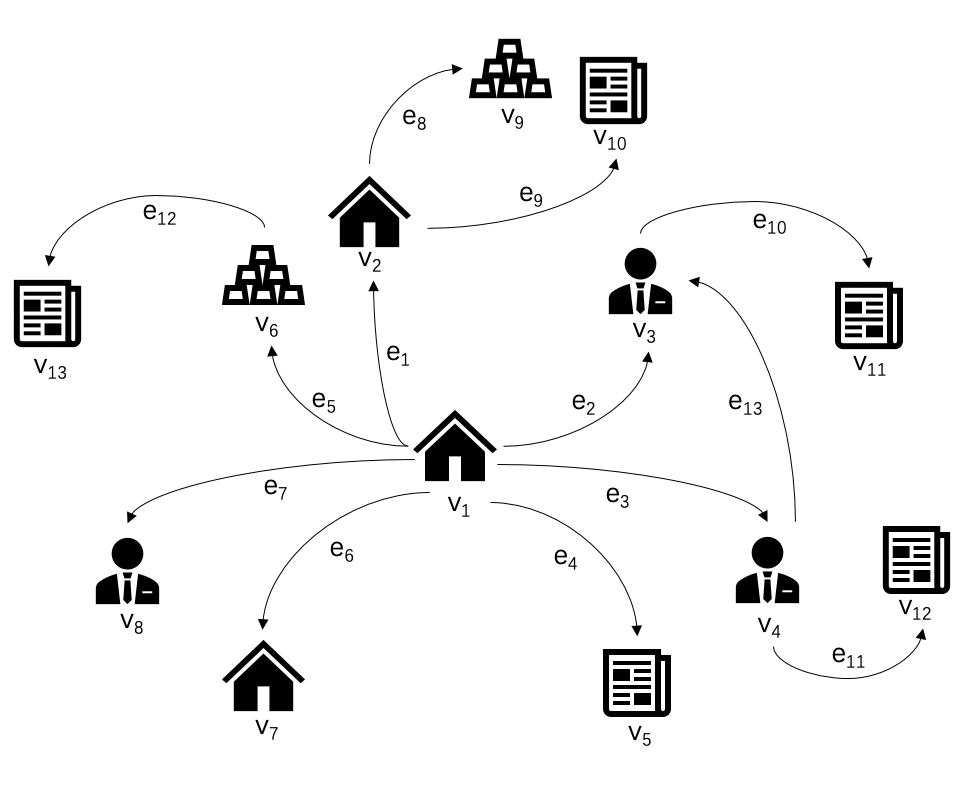}\newline
      \end{center}
             \caption{The SME heterogeneous information network}
             \label{fig:sme}
\end{figure}

With the SME schema defined, an example of SME heterogeneous information network is shown in Figure \ref{fig:sme}.
We can see that $v_{1}$, $v_{2}$, $v_{7}$ are \emph{enterprise}, that we have $\tau(v_{1})=\mathcal{A}_{e}$, the same as $\tau(v_{2})$ and $ \tau(v_{7})$ are.
The $v_{6}$, $v_{9}$ are \emph{commodities}, that we have $\tau(v_{6})=\mathcal{A}_{c}$, the same as $\tau(v_{9})$.
The $v_{5}$, $v_{10}$, $v_{11}$, $v_{12}$, $v_{13}$ are \emph{news}, that we have $\tau(v_{5})=\mathcal{A}_{n}$, the same as $\tau(v_{10})$, $\tau(v_{11})$, $\tau(v_{12})$, and $\tau(v_{13})$ are.
The $v_{3}$, $v_{4}$, $v_{8}$ are \emph{persons}, that we have $\tau(v_{3})=\mathcal{A}_{p}$, the same as $\tau(v_{4})$ and $\tau(v_{8})$ are.
The $e_{5}$, $e_{8}$ are the relation of produces, that we have $\phi(e_{5})=\mathcal{R}_{produce}$, the same as $\phi(e_{8})$. 
The $e_{4}$, $e_{9}$,$e_{10}$, $e_{11}$, $e_{12}$ are the relation of \emph{reports}, that we have $\phi(e_{4})=\mathcal{R}_{report}$, the same as $\phi(e_{9})$, $\phi(e_{10})$, $\phi(e_{11})$, and $\phi(e_{12})$ are. 
The $e_{6}$ is the relation of \emph{supply}, $e_{1}$ is the relation of \emph{parent}, that we have $\phi(e_{6})=\mathcal{R}_{supply}$, and $\phi(e_{1})=\mathcal{R}_{parent}$. 
The $e_{7}$, $e_{2}$ are the relation of \emph{controller}, $e_{3}$ is the relation of \emph{employee}, that we have $\phi(e_{7})=\mathcal{R}_{control}$, the same as $\phi(e_{2})$, and $\phi(e_{3})=\mathcal{R}_{employee}$. 
The $e_{13}$ is the relation of \emph{relate}, that we have $\phi(e_{13})=\mathcal{R}_{relate}$.

\subsection{SME Meta Path}\label{subsec:meta}

In the SME network graph we build in Section \ref{subsec:def}, a graph edge is used to present the relationship between two objects.
Limited by the definition of edge, the represented relationships can only be some simple ones, which are insufficient to describe the relationships used in the problem of SME credit risk.
In order to model complicated relationships, in this section, we introduce another data structure, meta path (MP), to represent complicated and implicit relations in SME network.

\begin{definition}\label{def:meta}
With a schema $S=(\mathcal{A},\mathcal{R})$, a \emph{meta path} $P$ is a path in the form  $\mathcal{A}_{1} \stackrel{ \mathcal{R}_{1} }{\longrightarrow} \mathcal{A}_{2} \stackrel{ \mathcal{R}_{2} }{\longrightarrow}...\stackrel{ \mathcal{R}_{n} }{\longrightarrow} \mathcal{A}_{ n+1}$ which defines a composite relation $\mathcal{R} = \mathcal{R}_{1} \circ \mathcal{R}_{2} \circ \ldots \circ \mathcal{R}_{n}$ between $\mathcal{A}_{1}$ and $\mathcal{A}_{n+1}$, where $\circ$ denotes the composition operator on relations.
\end{definition}

For simplicity, we use the names of object types and relation types denoting the MP: $P = \mathcal{A}_{1}{\cdot}\mathcal{R}_{1}{\cdot} \mathcal{A}_{2} \ldots \mathcal{R}_{n}{\cdot}\mathcal{A}_{n+1}$.
With the definition of meta path, a path $p = v_{1}{\cdot}e_{1}{\cdot} v_{2} \ldots e_{n}{\cdot}v_{n+1} $ in graph $G$ follows a meta path $P$, if for any vertex $v_{i}\in\mathcal{V}$ and any edge $e_{i} \in \mathcal{E}$, there have edge $e_{i}$ is between $v_{i}$ and $v_{i+1}$, $\tau(v_{i})={\mathcal A_{i}}$  and $\phi(e_{i})={\mathcal R_{i}}$.
We also call $p$ as a \emph{path instance} of $P$ with the denotation $p\in P$.

According to the definition, some examples of meta paths can be seen in Figure \ref{fig:schema}.
$P = \mathcal{A}_{e}{\cdot}\mathcal{R}_{parent}{\cdot}\mathcal{A}_{e}{\cdot}\mathcal{R}_{report}{\cdot}\mathcal{A}_{n}$ is a MP, which represent the information that  the SME's parent enterprise  has report a news.
According to Figure \ref{fig:sme}, there is a path instance $p = v_{1}{\cdot}e_{1}{\cdot} v_{2}{\cdot} e_{9}{\cdot}v_{10}$ of MP $P$.
Because $\tau(v_{1})=\mathcal{A}_{e}$, $\tau(v_{2})=\mathcal{A}_{e}$, $\tau(v_{10})=\mathcal{A}_{n}$, $\phi(e_{1})= \mathcal{R}_{parent}$, $\phi(e_{9})=\mathcal{R}_{report}$.

The given MP definition structures logical connections between objects, making our model more expressive and interpretable.
It not only can show explicit reasons of factors affecting SMEs on credit risk, but also can explain implicit logics of correlation between objects having no direct links in SME information network.

Compared to the information carried by objects, the information carried by meta paths is more critical in evaluating the credit risk of SMEs.
The reason is that the expression ability of meta paths is more stronger.
Through different meta paths, the same financial object may affect another financial object significantly differently.
For instance, in Figure \ref{fig:sme} we can see that there exist two paths from person $v_{4}$ to enterprise $v_{1}$.
The first one is $p = v_{1}{\cdot}e_{3}{\cdot} v_{4}$ following meta path  $P = \mathcal{A}_{e}{\cdot}\mathcal{R}_{employee}{\cdot} \mathcal{A}_{p}$ and the second one is $p = v_{1}{\cdot}e_{2}{\cdot} v_{3}{\cdot}e_{13}{\cdot} v_{4}$ following meta path $P = \mathcal{A}_{e}{\cdot}\mathcal{R}_{control}{\cdot} \mathcal{A}_{p}{\cdot}\mathcal{R}_{relate}{\cdot} \mathcal{A}_{p}$.
From the first path, the bribery scandal of an outsourcing employee $v_{4}$ may do limited harm to the enterprise $v_{1}$ since $v_{1}$ may have many other outsourcing employees to replace the role of $v_{4}$.
However, from the second path, the bribery scandal of the outsourcing employee $v_{4}$  may do significant harm to enterprise $v_{1}$ since $v_{4}$ has a domestic relation with $v_{3}$ who directs enterprise $v_{1}$.
Therefore, instead of inspecting each object's direct impact, our model regards a whole logical path consisting of objects and relations as a factor, in evaluating the credit risk of SMEs.

\section{Meta Path Impact On SME}\label{sec:impact}
In the above, we have given the definition of MP, a well-patterned structure to represent various semantics relating to SME credit risk.
It has been shown that even with no direct link given, the negative information of some SME may affect others heavily through meta paths.
For example, a piece of negative news about an enterprise director may lead to a bad reputation for his enterprise;
a low-quality product of a parent enterprise may cause a loss of competitiveness to its subsidiary enterprises.
Usually, potential risks brought from paths is non-trivial to be neglected when an SME is evaluated, but how to formulate such potential risk remains a question.
In order to solve this question, in this section, we will propose several novel features, named meta path feature, to represent the risk.

\subsection{Risk Inference from Object}\label{subsubsec:nega}
Before introducing meta path features, we first give a method to identify if there exists potential risk in financial objects themselves.
According to the object types studied in Section \ref{subsec:def}, except the\emph{news} object which is used to provide negative or positive information,
a \emph{commodity} object is regarded with potential risks if its quality is not reliable;
a \emph{person} object is regarded with potential risks if his capability is not qualified;
an \emph{enterprise} object is regarded with potential risks if it is lack of credibility.
In this paper, in order to inference if potential risks exist, considering applicability and generality, we use Naive Bayes model to inference if the mentioned objects is risky or not.
Our probabilistic model is learnt from public historical data, such as financial statements, annual reports, and online public news.
The definition of our Naive Bayes inference model is given as the following:
\begin{definition}\label{def:gamma}
    With the assumption that each attribute feature of an object is independent of each other, we define an inference function $\Gamma(x)$ to evaluate if object $x$ is risky based on the probability $\mathbb{P}(y=1|x)$ learned from Naive Bayes model.   
    \begin{scriptsize}
        \begin{equation}
            \begin{aligned}
              \Gamma(x) &=
                    \begin{cases}
                          1               & \mathbb{P} ( y=1|x )>0.5\\
                          0               & \text{otherwise}
                         \end{cases}\\
              \mathbb{P} ( y=1|x ) &= \frac  { { \prod_{i} ^ {n} } \mathbb{P} (x ^ {(i)}|y=1) \mathbb{P} (y=1)}  { { \prod_{i} ^ {n} } \mathbb{P} (x ^ {(i)} |y=1) \mathbb{P} (y=1) + { \prod_{i} ^ {n} } \mathbb{P} (x ^ {(i)}|y=0) \mathbb{P}(y=0) }
            \end{aligned}
        \end{equation}
    \end{scriptsize}
\\where $x^{(i)}$ is the $i$th attribute feature of object $x$, $n$ is the number of all attributes, $y=1$ indicates risky object and $y=0$ indicates non-risky object.
\end{definition}

With the inference function, we are able to identify the risk of a financial object by its own information.
For instance, a \emph{commodity} object with low sales volume, high repair rate, and high refund will be inferred as risky one; 
a \emph{person} object with irrelevant education background, irrelevant working experience, and short working years will be inferred as risky one;
an \emph{enterprise} object with low ROE ratio, low quick ratio, and high asset-liability ratio will be inferred as risky one.
In the next, we will study how to inference the potential risk from MP level.

\subsection{Risk Inference from Meta Path}\label{subsubsec:fea}
\begin{figure}[pth]
      \begin{center}
             \includegraphics[width=2in]{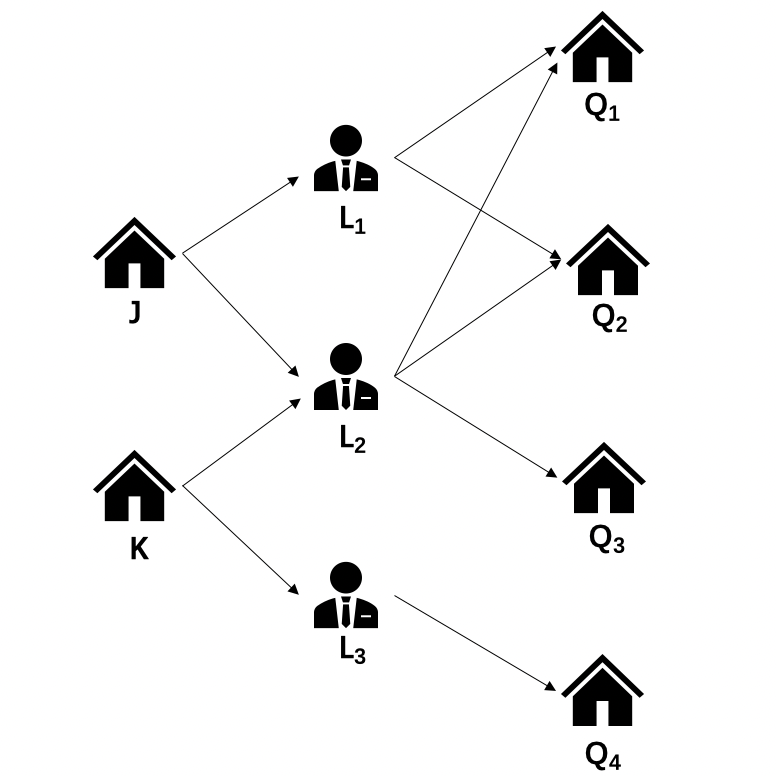}\newline
      \end{center}
             \caption{The path instances of MP $P = \mathcal{A}_{e}{\cdot}\mathcal{R}_{control}{\cdot}\mathcal{A}_{p}{\cdot}\mathcal{R}_{shareholder}{\cdot}\mathcal{A}_{e}$. $J$ and $K$ are target SMEs, $L_{1}$,  $L_{2}$ and $L_{3}$ are controllers of $J$ and $K$.  $Q_{1}$,  $Q_{2}$, $Q_{3}$ and $Q_{4}$ are the associated enterprises of controllers $L_{1}$,  $L_{2}$ and $L_{3}$}.
             \label{fig:instance}
\end{figure}

In an SME information network, an enterprise may have many paths linking to other financial objects, as shown in Figure \ref{fig:instance}.
We can see enterprise \emph{$J$} has $5$ path instances for meta path $P$ = $\mathcal{A}_{e}{\cdot}\mathcal{R}_{control}{\cdot}\mathcal{A}_{p}{\cdot}\mathcal{R}_{shareholder}{\cdot} \mathcal{A}_{e}$ and enterprise \emph{$K$} has $4$ path instances for MP $P$.

With the inference function defined above, we are able to identify if objects in the above information network are risky or not.
Thus, for a specific MP, with the objects linked by its path instances, it is natural to infer that an enterprise is most likely to be risky if potential risks exist in most of its linked objects.
Based on this straight intuition, we next present several features to elaborate such risk from meta paths.

\subsubsection{Meta Path Feature}\label{subsubsec:fea}
Given an enterprise $x$, the number of risky objects connected by a MP $P$ are taken as an indicator to reflect the impact of meta path $P$ on target enterprise $x$.
The larger the indicator is, the higher the potential risk exists.
Formally, we call the indicator as naive MP feature and give its definition as the following.

\begin{definition}\label{def:ind}
    \emph{Naive MP feature $N_P(x)$} is an indicator to reveal the impact of meta path $P$ on enterprise $x$: 
    \begin{scriptsize}
        \begin{equation}
            \begin{aligned}
              N_P(x)= \frac { | \{ x' \in D | \exists p_{x \rightsquigarrow x'} \in P,\Gamma(x')=1 \} |}  { | \{ x' \in D |  \exists p_{x \rightsquigarrow x'} \in P \} |} \\
             \end{aligned}
        \end{equation}
    \end{scriptsize}
\\where $D$ is an SME object collection, $p_{ x\rightsquigarrow x' }$ is a path instance from object $x$ to object $x'$ and  $\Gamma(x)$ is the inference function defined in Section \ref{subsubsec:nega}.
\end{definition}

In Figure \ref{fig:instance}, if $Q_{2}$, $Q_{3}$ and $Q_{4}$ are risky objects,
then we have $N_P(J)$ = $3/5$ = $0.6$, $N_P(K)$ = $3/4$ = $0.75$.

\subsubsection{Weighted Meta Path Feature}\label{subsubsec:weightedfea}
Although the above meta path feature can effectively indicate the impact of MP, it may be argued that the impact of different objects on the same MP should not be the same.
For all the objects in the network, irrelevant objects may affect small; relevant ones may matter big.
Especially for an SME, the enterprise, which is its parent company, should influence it deeper than the enterprise, which only has one cooperation with it.
Therefore, instead of treating all objects equally, it is more reasonable to treat them differently according to their relevance with the target SME.
Next, considering relevance between objects, we will give a relevance-weighted version of meta path feature accordingly.

Usually, relevance is used to measure how close two objects distance to each other.
As there is no unified definition of relevance, different applications have unique and appropriate relevance measures.
In SME application, there exists a usual fact that, even though an enterprise is of well financial status, it may also default, which is caused by the propagated negative influence of its related upstream and downstream enterprises.
Therefore, to measure the relevance between SME objects, a logical structure-based relevance measure is better than a textual context-based relevance measure.

A straightforward idea is that for any object pair, the two which have more paths should be more relevant.
From this idea, we simply introduce a path count version of MP weighted feature as follows.  
\begin{definition}\label{def:npc}
    \emph{CountSim MP weight feature} $C_P(x)$ is an indicator to reveal the structure relevance impact of meta path P on enterprise $x$. We call it \emph{CountSim MP feature}.
    \begin{scriptsize}
        \begin{equation}
            \begin{aligned}
              C_P(x) = \frac{ | \{ x' \in D |  \exists p_{x \rightsquigarrow x'} \in P \} |}  {{{ | \{ x \in S\} } |}+{ | \{ x' \in S' \}} |} \\
             \end{aligned}
        \end{equation}
    \end{scriptsize}
\\where $S$ and $S'$ are SME object collections where all links from $x$ and to $x'$ respectively. $D$ is another SME object collection where contains all objects.
\end{definition}

The path count version is simple to apply but it make little use of graph structure.
In SME heterogeneous information network, logical relationships between objects are captured by the structure of graph paths.
Hence, compared to other measures, a path-based measure of relevance is more appropriate to be adopted in our model.
At last, we apply HeteSim \cite{shi2014hetesim}, an effective path-based similarity, to evaluate the relevance between objects.

\begin{definition}    
    \emph{HeteSim MP weight feature $H_P(x)$ takes HeteSim as the similarity measure to reveal the path relevance impact of meta path P on enterprise $x$. We call it \emph{HeteSim MP feature}. }\\
    \begin{scriptsize}
        \begin{equation}
            \begin{aligned}
              H_P(x)=  \frac { { \sum\nolimits_{ x' \in  \{ x'| \exists p_{x \rightsquigarrow x'} \in P, \Gamma(x')=1 \} } } HeteSim(x,x') }  { { \sum\nolimits_{ x' \in  \{ x'| \exists p_{x \rightsquigarrow x'} \in P \} } } HeteSim(x,x') }
            \end{aligned}
        \end{equation}
    \end{scriptsize}
\\where $p_{ x\rightsquigarrow x' }$ is a path instance from object $x$ to object $x'$,  $HeteSim(x,x')$ is the relevance between object $x$ and object $x'$ under HeteSim and  $\Gamma(x)$ is the estimating function defined in Section \ref{subsubsec:nega}.
\end{definition}

\section{Experiments}\label{sec:experiments}
In this section, we are going to investigate the effectiveness of meta path features.
We conduct experiments on three real-world SME datasets.
The result and explanation is detailed in this part.

\subsection{Data and settings}\label{subsec:dataset}
In our experiments, three datasets recording enterprises' statistics are used for comparison.
GEM(The Growth Enterprise Market from Shenzhen Stock Exchange) and STAR (The Science and Technology Innovation Board from Shanghai Stock Exchange) datasets are about the SMEs of high technology, and SB (The Small and Medium-sized Enterprise Board from Shenzhen Stock Exchange) dataset are about traditional enterprises.
All the datasets can be downloaded from CSMAR (\url{https://www.gtarsc.com}).
As this paper only considers four types of financial entities (\emph{person}, \emph{commodity}, \emph{enterprise} and \emph{news}), our experiments are only performed on the enterprises those at least relate to one person, one commodity, one other enterprise and one piece of news. 

The risk information about whether an enterprise is lack of credibilities, a person is lack of qualifications and a commodity is lack of reliabilities is obtained from CSMAR and cninf (\url{http://www.cninfo.com.cn}), which provide an authoritative and professional assessment on the entities. 
The news information is collected from China Judgements Online (\url{https://wenshu.court.gov.cn}).
The final details of datasets is shown in Table \ref{tab:statistical}.
\begin{table}
\caption{Dataset information}
\label{tab:statistical}       
\centering
\begin{tabular}{lccc}
  \hline\noalign{\smallskip}
      & GEM & STAR & SB\\
     \noalign{\smallskip}\hline\noalign{\smallskip}
     number of enterprise & 528&297& 722\\
     related enterprise information & 58478&26554& 80729\\
      related person information & 360462 &38663& 515504\\
       related news information & 13026&3748&24718  \\
        related commodities information & 17450&8987 & 36735\\  
  \noalign{\smallskip}\hline
\end{tabular}
\end{table}
As the gathered risk information may not be complete, for some important but unknown entities, we use the model in Section \ref{subsubsec:nega} to infer their risk.
If an entity's inferred probability is larger than $0.75$, it is deemed as risky. 

Since the brought impact from a meta path decreases with its length increasing, we only consider the meta paths with length less than 6.
And, the meta paths which do not start with SME type are not selected for our experiments.
With the proposed MP features, we test their performance using a default prediction model which is used to learn the weights associated with those features.
The logistic regression model is taken as the prediction model, which is optimized by MLE (Maximum Likelihood Estimation).

In this section, all experiments were performed using Python 2.7.17 in Win $8.1+$ with CPU $i5-9300+$ processor and $8G+$ RAM.

\subsection{Selection of Meta Path Features}\label{subsec: mp feature selection}
Even limited by the length constraint, there may still exist numerous meta paths.
Among all possible meta path features, which ones are the most valuable ones?
In this section, we will run experiments to show the importance of meta path features.

We first generate 40 meta path features according to Definition \ref{def:ind} for simplicity.
Then each feature is tested under Wald test and the $p$-value of the feature associated with its meta path is used to evaluate the feature's importance.
The test is performed on all three datasets.
From the results, we list the top 20 significant meta path features for each dataset in Tables \ref{tab:1} - \ref{tab:SMEM} and the bottom 20 meta path features in Table \ref{tab:BGEMM} - \ref{tab:BSMEM} .
We can see that for all three datasets, the controller's ability, parent enterprise financial status, and news reported for enterprise, plays very significant roles in determining SME status.
However, the longer the relation chains, the worse the performance of MP features.
This may be due to the fact that longer links contain less valuable information.
Or the longer the chain, the more distracting and inaccurate information it contains.
Look into details, we find that for GEM and STAR datasets, the MP features containing personnel relations are most significant, while those containing enterprise relations are the least.
For SB dataset, the opposite is true.
It is reasonable, that the conventional SME, due to their own resource constraints, will pay more attention to the relationship with stakeholders in order to ensure stable development.
The high-technology SME mainly focus on technology research and development, so the ability of personnel has a significant impact on the enterprise.

\begin{table}
\caption{Top -$20$ significant meta path features for the GEM dataset}
\label{tab:1}       
\centering
\begin{threeparttable}
\begin{tabular}{lllll}
\hline\noalign{\smallskip}
 &  Meta path feature & P-value &Significance level\tnote{1}\\
\hline\noalign{\smallskip}
                   1&$\mathcal{A}_{e}\cdot\mathcal{R}_{control}\cdot\mathcal{A}_{p}$ & \textbf{3.7876e-46}&**** \\
                   2&$\mathcal{A}_{e}\cdot\mathcal{R}_{parent}\cdot\mathcal{A}_{e}$  &\textbf{5.3500e-37}&****\\
                   3&$\mathcal{A}_{e}\cdot\mathcal{R}_{report}\cdot\mathcal{A}_{n}$&\textbf{1.7758e-32}&****\\
                   4&$\mathcal{A}_{e}\cdot\mathcal{R}_{control}\cdot\mathcal{A}_{p}\cdot\mathcal{R}_{control}\cdot\mathcal{A}_{e} $&1.0156e-32&****\\
                   5&$\mathcal{A}_{e}\cdot\mathcal{R}_{produce}\cdot\mathcal{A}_{c}\cdot\mathcal{R}_{report}\cdot\mathcal{A}_{n}$&3.9645e-29&****\\
                   6&$\mathcal{A}_{e}\cdot\mathcal{R}_{manager}\cdot\mathcal{A}_{p}$&8.3629e-26&****\\
                   7&$\mathcal{A}_{e}\cdot\mathcal{R}_{produce}\cdot\mathcal{A}_{c}$&2.2358e-26&****\\
                   8&$\mathcal{A}_{e}\cdot\mathcal{R}_{boardmember}\cdot\mathcal{A}_{p}$&6.1598e-23&****\\
                   9&$\mathcal{A}_{e}\cdot\mathcal{R}_{shareholder}\cdot\mathcal{A}_{p}$&2.4664e-15&****\\
                  10&$\mathcal{A}_{e} \cdot\mathcal{R}_{control}\cdot\mathcal{A}_{p}\cdot\mathcal{R}_{report}\cdot\mathcal{A}_{n}$&1.6067e-9&****\\
                  11&$\mathcal{A}_{e}\cdot\mathcal{R}_{parent}\cdot\mathcal{A}_{e}\cdot\mathcal{R}_{manager}\cdot\mathcal{A}_{p}$ & 3.7876e-6&**** \\
                  12&$\mathcal{A}_{e}\cdot\mathcal{R}_{subsidiary}\cdot\mathcal{A}_{e} $&5.3500e-5&****\\
                  13&$\mathcal{A}_{e}\cdot\mathcal{R}_{manager}\cdot\mathcal{A}_{p}\cdot\mathcal{R}_{report}\cdot\mathcal{A}_{n}$&0.00121&***\\
                  14&$\mathcal{A}_{e}\cdot\mathcal{R}_{control}\cdot\mathcal{A}_{p}\cdot\mathcal{R}_{relate}\cdot\mathcal{A}_{p}$&0.00160&***\\
                  15&$\mathcal{A}_{e}\cdot\mathcal{R}_{subsidiary}\cdot\mathcal{A}_{e}\cdot\mathcal{R}_{report}\cdot\mathcal{A}_{n}$&0.00236&***\\
                  16&$\mathcal{A}_{e}\cdot\mathcal{R}_{control}\cdot\mathcal{A}_{p}\cdot\mathcal{R}_{manager}\cdot\mathcal{A}_{e} $&0.00246 &***\\
                  17&$\mathcal{A}_{e}\cdot\mathcal{R}_{subsidiary}\cdot\mathcal{A}_{e}\cdot\mathcal{R}_{control}\cdot\mathcal{A}_{p}$&0.00396&***\\
                  18&$\mathcal{A}_{e}\cdot\mathcal{R}_{parent}\cdot\mathcal{A}_{e}\cdot\mathcal{R}_{report}\cdot\mathcal{A}_{n}$&0.00615&***\\
                  19&$\mathcal{A}_{e}\cdot\mathcal{R}_{parent}\cdot\mathcal{A}_{e}\cdot\mathcal{R}_{control}\cdot\mathcal{A}_{p}$&0.00758&***\\
                   20&$\mathcal{A}_{e}\cdot\mathcal{R}_{supply}\cdot\mathcal{A}_{e}$& 0.00823&***\\
\noalign{\smallskip}\hline
\end{tabular}
\begin{tablenotes}
 \footnotesize
 \item[1] *: p\textless 0.1, **: p\textless 0.05, ***: p\textless0.01, ****: p\textless 0.001
\end{tablenotes}
\end{threeparttable}
\end{table}

\begin{table}
\caption{Top -$20$ significant meta path features for the STAR dataset}
\label{tab:STARM}       
\centering
\begin{threeparttable}
\begin{tabular}{lllll}
\hline\noalign{\smallskip}
 &  Meta path feature & P-value&Significance level\tnote{1} \\
\hline\noalign{\smallskip}
                    1&$\mathcal{A}_{e}\cdot\mathcal{R}_{control}\cdot\mathcal{A}_{p}$ & \textbf{7.4107e-44}&**** \\
                    2&$\mathcal{A}_{e}\cdot\mathcal{R}_{parent}\cdot\mathcal{A}_{e}$  &\textbf{3.3610e-37}&****\\
                    3&$\mathcal{A}_{e}\cdot\mathcal{R}_{shareholder}\cdot\mathcal{A}_{p}$&\textbf{1.8247e-29}&****\\
                    4&$\mathcal{A}_{e}\cdot\mathcal{R}_{report}\cdot\mathcal{A}_{n} $&1.8709e-22&****\\
                    5&$\mathcal{A}_{e}\cdot\mathcal{R}_{subsidiary}\cdot\mathcal{A}_{e}$&1.925e-17&****\\
                    6&$\mathcal{A}_{e}\cdot\mathcal{R}_{manager}\cdot\mathcal{A}_{p}$&2.7723e-11&****\\
                    7&$\mathcal{A}_{e}\cdot\mathcal{R}_{boardmember}\cdot\mathcal{A}_{p}$&9.2910e-8&****\\
                    8&$\mathcal{A}_{e}\cdot\mathcal{R}_{control}\cdot\mathcal{A}_{p}\cdot\mathcal{R}_{report}\cdot\mathcal{A}_{n} $&2.8380e-4&****\\
                    9&$\mathcal{A}_{e}\cdot\mathcal{R}_{subsidiary}\cdot\mathcal{A}_{e}\cdot\mathcal{R}_{report}\cdot\mathcal{A}_{n}$&0.000929&****\\
                   10&$\mathcal{A}_{e}\cdot\mathcal{R}_{produce}\cdot\mathcal{A}_{c}$&0.00175&***\\
                   11&$\mathcal{A}_{e}\cdot\mathcal{R}_{control}\cdot\mathcal{A}_{p}\cdot\mathcal{R}_{control}\cdot\mathcal{A}_{e}$ & 0.00277&*** \\
                   12&$\mathcal{A}_{e}\cdot\mathcal{R}_{produce}\cdot\mathcal{A}_{c}\cdot\mathcal{R}_{report}\cdot\mathcal{A}_{n}$  &0.00283&***\\
                   13&$\mathcal{A}_{e}\cdot\mathcal{R}_{boardmember}\cdot\mathcal{A}_{p}\cdot\mathcal{R}_{report}\cdot\mathcal{A}_{n}$&0.00341&***\\
                   14&$\mathcal{A}_{e}\cdot\mathcal{R}_{supply}\cdot\mathcal{A}_{p}$&0.0044&***\\
                   15&$\mathcal{A}_{e}\cdot\mathcal{R}_{parent}\cdot\mathcal{A}_{e}\cdot\mathcal{R}_{control}\cdot\mathcal{A}_{p}$&0.00455&***\\
                   16&$\mathcal{A}_{e}\cdot\mathcal{R}_{sales}\cdot\mathcal{A}_{e} $&0.00476&***\\
                   17&$\mathcal{A}_{e}\cdot\mathcal{R}_{parent}\cdot\mathcal{A}_{e}\cdot\mathcal{R}_{manager}\cdot\mathcal{A}_{p}$&0.00496&***\\
                   18&$\mathcal{A}_{e}\cdot\mathcal{R}_{manager}\cdot\mathcal{A}_{p}\cdot\mathcal{R}_{report}\cdot\mathcal{A}_{n}$&0.00510&***\\
                   19&$\mathcal{A}_{e}\cdot\mathcal{R}_{supply}\cdot\mathcal{A}_{e}\cdot\mathcal{R}_{report}\cdot\mathcal{A}_{n}$&0.00528&***\\
                   20&$\mathcal{A}_{e}\cdot\mathcal{R}_{parent}\cdot\mathcal{A}_{e}\cdot\mathcal{R}_{report}\cdot\mathcal{A}_{n}$&0.00741&***\\
\noalign{\smallskip}\hline
\end{tabular}
\begin{tablenotes}
 \footnotesize
 \item[1] *: p\textless 0.1, **: p\textless 0.05, ***: p\textless0.01, ****: p\textless 0.001
\end{tablenotes}
\end{threeparttable}
\end{table}

\begin{table}
\caption{Top -$20$ significant meta path features for the SB dataset}
\label{tab:SMEM}       
\centering
\begin{threeparttable}
\begin{tabular}{lllll}
\hline\noalign{\smallskip}
 &  Meta path feature & P-value&Significance level\tnote{1} \\
\hline\noalign{\smallskip}
                    1&$\mathcal{A}_{e}\cdot\mathcal{R}_{report}\cdot\mathcal{A}_{n}$ & \textbf{1.2831e-48}&**** \\
                    2&$\mathcal{A}_{e}\cdot\mathcal{R}_{parent}\cdot\mathcal{A}_{e}$  &\textbf{3.0306e-45}&****\\
                    3&$\mathcal{A}_{e}\cdot\mathcal{R}_{control}\cdot\mathcal{A}_{p} $&\textbf{1.5510e-36}&****\\
                    4&$\mathcal{A}_{e}\cdot\mathcal{R}_{subsidiary}\cdot\mathcal{A}_{e}$&6.5260e-35&****\\
                    5&$\mathcal{A}_{e}\cdot\mathcal{R}_{subsidiary}\cdot\mathcal{A}_{e}\cdot\mathcal{R}_{report}\cdot\mathcal{A}_{n}$&3.7263e-35&****\\
                    6&$\mathcal{A}_{e}\cdot\mathcal{R}_{control}\cdot\mathcal{A}_{p}\cdot\mathcal{R}_{control}\cdot\mathcal{A}_{e}$&4.4973e-33&****\\
                    7&$\mathcal{A}_{e}\cdot\mathcal{R}_{control}\cdot\mathcal{A}_{p}\cdot\mathcal{R}_{manager}\cdot\mathcal{A}_{e}$&2.3524e-33&****\\
                    8&$\mathcal{A}_{e}\cdot\mathcal{R}_{supply}\cdot\mathcal{A}_{e}$&1.1475e-28&****\\
                    9&$\mathcal{A}_{e}\cdot\mathcal{R}_{boardmember}\cdot\mathcal{A}_{p}$&6.8367e-27&****\\
                   10&$\mathcal{A}_{e}\cdot\mathcal{R}_{parent}\cdot\mathcal{A}_{e}\cdot\mathcal{R}_{manager}\cdot\mathcal{A}_{p}$&5.2674e-13&****\\
                   11&$\mathcal{A}_{e}\cdot\mathcal{R}_{produce}\cdot\mathcal{A}_{c}$ & 1.2831e-11&**** \\
                   12&$\mathcal{A}_{e}\cdot\mathcal{R}_{boardmember}\cdot\mathcal{A}_{p}\cdot\mathcal{R}_{report}\cdot\mathcal{A}_{n}$  &3.0306e-9&****\\
                   13&$\mathcal{A}_{e}\cdot\mathcal{R}_{shareholder}\cdot\mathcal{A}_{p} $&1.5510e-8&****\\
                   14&$\mathcal{A}_{e}\cdot\mathcal{R}_{control}\cdot\mathcal{A}_{p}\cdot\mathcal{R}_{shareholder}\cdot\mathcal{A}_{e}$&6.5260e-6&****\\
                   15&$\mathcal{A}_{e}\cdot\mathcal{R}_{sales}\cdot\mathcal{A}_{e}$&3.7263e-5&****\\
                   16&$\mathcal{A}_{e}\cdot\mathcal{R}_{parent}\cdot\mathcal{A}_{e}\cdot\mathcal{R}_{produce}\cdot\mathcal{A}_{p}$&4.4973e-4&****\\
                   17&$\mathcal{A}_{e}\cdot\mathcal{R}_{manager}\cdot\mathcal{A}_{p}\cdot\mathcal{R}_{control}\cdot\mathcal{A}_{e}$&2.3524e-4&****\\
                   18&$\mathcal{A}_{e}\cdot\mathcal{R}_{sales}\cdot\mathcal{A}_{e}\cdot\mathcal{R}_{report}\cdot\mathcal{A}_{n}$&0.00114&***\\
                   19&$\mathcal{A}_{e}\cdot\mathcal{R}_{parent}\cdot\mathcal{A}_{e}\cdot\mathcal{R}_{report}\cdot\mathcal{A}_{n}$&0.00526&***\\
                   20&$\mathcal{A}_{e}\cdot\mathcal{R}_{supply}\cdot\mathcal{A}_{e}\cdot\mathcal{R}_{report}\cdot\mathcal{A}_{n}$& 0.00683 &***\\
\noalign{\smallskip}\hline
\end{tabular}
\begin{tablenotes}
 \footnotesize
 \item[1] *: p\textless 0.1, **: p\textless 0.05, ***: p\textless0.01, ****: p\textless 0.001
\end{tablenotes}
\end{threeparttable}
\end{table}

\begin{table}
\caption{Bottom -$20$ significant meta path features for the GEM dataset}
\label{tab:BGEMM}       
\centering
\begin{threeparttable}
\begin{tabular}{lllll}
\hline\noalign{\smallskip}
 &  Meta path feature & P-value&Significance level\tnote{1} \\
\hline\noalign{\smallskip}
                   1&$\mathcal{A}_{e}\cdot\mathcal{R}_{sales}\cdot\mathcal{A}_{e}$ & 0.0783&* \\
                   2&$\mathcal{A}_{e}\cdot\mathcal{R}_{supply}\cdot\mathcal{A}_{e}\cdot\mathcal{R}_{report}\cdot\mathcal{A}_{n}$  &0.0778&*\\
                   3&$\mathcal{A}_{e}\cdot\mathcal{R}_{control}\cdot\mathcal{A}_{p}\cdot\mathcal{R}_{shareholder}\cdot\mathcal{A}_{e}$&0.0788&*\\
                   4&$\mathcal{A}_{e}\cdot\mathcal{R}_{shareholder}\cdot\mathcal{A}_{p}\cdot\mathcal{R}_{control}\cdot\mathcal{A}_{e} $&0.0832&*\\
                   5&$\mathcal{A}_{e}\cdot\mathcal{R}_{shareholder}\cdot\mathcal{A}_{p}\cdot\mathcal{R}_{shareholder}\cdot\mathcal{A}_{e}$&0.0854&*\\
                   6&$\mathcal{A}_{e}\cdot\mathcal{R}_{shareholder}\cdot\mathcal{A}_{p}\cdot\mathcal{R}_{report}\cdot\mathcal{A}_{n} $&0.0861&*\\
                   7&$\mathcal{A}_{e}\cdot\mathcal{R}_{sales}\cdot\mathcal{A}_{e}\cdot\mathcal{R}_{report}\cdot\mathcal{A}_{n}$&0.0874&*\\
                   8&$\mathcal{A}_{e}\cdot\mathcal{R}_{shareholder}\cdot\mathcal{A}_{p}\cdot\mathcal{R}_{manager}\cdot\mathcal{A}_{e}$&0.0889&*\\
                   9&$\mathcal{A}_{e}\cdot\mathcal{R}_{supply}\cdot\mathcal{A}_{e}\cdot\mathcal{R}_{produce}\cdot\mathcal{A}_{c}$&0.0893&*\\
                  10&$\mathcal{A}_{e}\cdot\mathcal{R}_{manager}\cdot\mathcal{A}_{p}\cdot\mathcal{R}_{control}\cdot\mathcal{A}_{e}$&0.0896&*\\
                  11&$\mathcal{A}_{e}\cdot\mathcal{R}_{sales}\cdot\mathcal{A}_{e}\cdot\mathcal{R}_{produce}\cdot\mathcal{A}_{c}$ & 0.0899&* \\
                  12&$\mathcal{A}_{e}\cdot\mathcal{R}_{parent}\cdot\mathcal{A}_{e}\cdot\mathcal{R}_{produce}\cdot\mathcal{A}_{c}$  &0.0932&*\\
                  13&$\mathcal{A}_{e}\cdot\mathcal{R}_{supply}\cdot\mathcal{A}_{e}\cdot\mathcal{R}_{manager}\cdot\mathcal{A}_{p}$&0.1775&-\\
                  14&$\mathcal{A}_{e}\cdot\mathcal{R}_{supply}\cdot\mathcal{A}_{e}\cdot\mathcal{R}_{control}\cdot\mathcal{A}_{p} $&2.4662&-\\
                  15&$\mathcal{A}_{e}\cdot\mathcal{R}_{control}\cdot\mathcal{A}_{p}\cdot\mathcal{R}_{employee}\cdot\mathcal{A}_{p}$&3.9645&-\\
                  16&$\mathcal{A}_{e}\cdot\mathcal{R}_{sales}\cdot\mathcal{A}_{e}\cdot\mathcal{R}_{shareholder}\cdot\mathcal{A}_{p} $&6.1598&-\\
                  17&$\mathcal{A}_{e}\cdot\mathcal{R}_{manager}\cdot\mathcal{A}_{e}\cdot\mathcal{R}_{shareholder}\cdot\mathcal{A}_{p}$&7.4662&-\\
                  18&$\mathcal{A}_{e}\cdot\mathcal{R}_{sales}\cdot\mathcal{A}_{e}\cdot\mathcal{R}_{control}\cdot\mathcal{A}_{p}$&10.6710&-\\
                  19&$\mathcal{A}_{e}\cdot\mathcal{R}_{shareholder}\cdot\mathcal{A}_{p}\cdot\mathcal{R}_{employee}\cdot\mathcal{A}_{e}$&12.4639&-\\
                   20&$\mathcal{A}_{e}\cdot\mathcal{R}_{employee}\cdot\mathcal{A}_{p}\cdot\mathcal{R}_{manager}\cdot\mathcal{A}_{p}$&16.0762&-\\
\noalign{\smallskip}\hline
\end{tabular}
\begin{tablenotes}
 \footnotesize
 \item[1] *: p\textless 0.1, **: p\textless 0.05, ***: p\textless0.01, ****: p\textless 0.001
\end{tablenotes}
\end{threeparttable}
\end{table}

\begin{table}
\caption{Bottom -$20$ significant meta path features for the STAR dataset}
\label{tab:BSTARM}       
\centering
\begin{threeparttable}
\begin{tabular}{lllll}
\hline\noalign{\smallskip}
 &  Meta path feature& P-value&Significance level\tnote{1} \\
\hline\noalign{\smallskip}
                    1&$\mathcal{A}_{e}\cdot\mathcal{R}_{shareholder}\cdot\mathcal{A}_{p}\cdot\mathcal{R}_{relate}\cdot\mathcal{A}_{p}$ & 0.0538&* \\
                    2&$\mathcal{A}_{e}\cdot\mathcal{R}_{control}\cdot\mathcal{A}_{p}\cdot\mathcal{R}_{manager}\cdot\mathcal{A}_{e}$  &0.0598&*\\
                    3&$\mathcal{A}_{e}\cdot\mathcal{R}_{control}\cdot\mathcal{A}_{p}\cdot\mathcal{R}_{shareholder}\cdot\mathcal{A}_{e}$&0.0641&*\\
                    4&$\mathcal{A}_{e}\cdot\mathcal{R}_{shareholder}\cdot\mathcal{A}_{p}\cdot\mathcal{R}_{report}\cdot\mathcal{A}_{n} $&0.0870&*\\
                    5&$\mathcal{A}_{e}\cdot\mathcal{R}_{subsidiary}\cdot\mathcal{A}_{e}\cdot\mathcal{R}_{control}\cdot\mathcal{A}_{p}$&0.0873&*\\
                    6&$\mathcal{A}_{e}\cdot\mathcal{R}_{shareholder}\cdot\mathcal{A}_{p}\cdot\mathcal{R}_{shareholder}\cdot\mathcal{A}_{e} $&0.0881&*\\
                    7&$\mathcal{A}_{e}\cdot\mathcal{R}_{manager}\cdot\mathcal{A}_{p}\cdot\mathcal{R}_{control}\cdot\mathcal{A}_{p}$&0.0886&*\\
                    8&$\mathcal{A}_{e}\cdot\mathcal{R}_{parent}\cdot\mathcal{A}_{e}\cdot\mathcal{R}_{produce}\cdot\mathcal{A}_{c}$&0.0928&*\\
                    9&$\mathcal{A}_{e}\cdot\mathcal{R}_{subsidiary}\cdot\mathcal{A}_{e}\cdot\mathcal{R}_{produce}\cdot\mathcal{A}_{c}$&0.0941&*\\
                   10&$\mathcal{A}_{e}\cdot\mathcal{R}_{shareholder}\cdot\mathcal{A}_{p}\cdot\mathcal{R}_{manager}\cdot\mathcal{A}_{e}$&0.0951&*\\
                   11&$\mathcal{A}_{e}\cdot\mathcal{R}_{subsidiary}\cdot\mathcal{A}_{e}\cdot\mathcal{R}_{manager}\cdot\mathcal{A}_{p}$ &0.0974&* \\
                   12&$\mathcal{A}_{e}\cdot\mathcal{R}_{supply}\cdot\mathcal{A}_{e}\cdot\mathcal{R}_{control}\cdot\mathcal{A}_{p}$  &0.0976&*\\
                   13&$\mathcal{A}_{e}\cdot\mathcal{R}_{sales}\cdot\mathcal{A}_{e}\cdot\mathcal{R}_{report}\cdot\mathcal{A}_{n}$&0.0982&*\\
                   14&$\mathcal{A}_{e}\cdot\mathcal{R}_{sales}\cdot\mathcal{A}_{e}\cdot\mathcal{R}_{produce}\cdot\mathcal{A}_{c} $&0.0987&*\\
                   15&$\mathcal{A}_{e}\cdot\mathcal{R}_{control}\cdot\mathcal{A}_{p}\cdot\mathcal{R}_{employee}\cdot\mathcal{A}_{e}$&4.6731&-\\
                   16&$\mathcal{A}_{e}\cdot\mathcal{R}_{sales}\cdot\mathcal{A}_{e}\cdot\mathcal{R}_{control}\cdot\mathcal{A}_{p} $&7.7232&-\\
                   17&$\mathcal{A}_{e}\cdot\mathcal{R}_{supply}\cdot\mathcal{A}_{e}\cdot\mathcal{R}_{produce}\cdot\mathcal{A}_{c}$&9.2910&-\\
                   18&$\mathcal{A}_{e}\cdot\mathcal{R}_{shareholder}\cdot\mathcal{A}_{p}\cdot\mathcal{R}_{control}\cdot\mathcal{A}_{e}$&12.8380&-\\
                   19&$\mathcal{A}_{e}\cdot\mathcal{R}_{supply}\cdot\mathcal{A}_{e}\cdot\mathcal{R}_{manager}\cdot\mathcal{A}_{p}$&14.4176&-\\
                   20&$\mathcal{A}_{e}\cdot\mathcal{R}_{shareholeder}\cdot\mathcal{A}_{p}\cdot\mathcal{R}_{employee}\cdot\mathcal{A}_{e}$&17.5919&-\\
\noalign{\smallskip}\hline
\end{tabular}
\begin{tablenotes}
 \footnotesize
 \item[1] *: p\textless 0.1, **: p\textless 0.05, ***: p\textless0.01, ****: p\textless 0.001
\end{tablenotes}
\end{threeparttable}
\end{table}

\begin{table}
\caption{Bottom -$20$ significant meta path features for the SB dataset}
\label{tab:BSMEM}       
\centering
\begin{threeparttable}
\begin{tabular}{lllll}
\hline\noalign{\smallskip}
 &  Meta path feature & P-value&Significance level\tnote{1} \\
\hline\noalign{\smallskip}
                    1&$\mathcal{A}_{e}\cdot\mathcal{R}_{produce}\cdot\mathcal{A}_{c}\cdot\mathcal{R}_{report}\cdot\mathcal{A}_{n}$ & 0.0714&* \\
                    2&$\mathcal{A}_{e}\cdot\mathcal{R}_{manager}\cdot\mathcal{A}_{p}\cdot\mathcal{R}_{report}\cdot\mathcal{A}_{n}$  &0.0730&*\\
                    3&$\mathcal{A}_{e}\cdot\mathcal{R}_{shareholder}\cdot\mathcal{A}_{p}\cdot\mathcal{R}_{relate}\cdot\mathcal{A}_{p} $&0.07551&*\\
                    4&$\mathcal{A}_{e}\cdot\mathcal{R}_{parent}\cdot\mathcal{A}_{e}\cdot\mathcal{R}_{control}\cdot\mathcal{A}_{p}$&0.07652&*\\
                    5&$\mathcal{A}_{e}\cdot\mathcal{R}_{parent}\cdot\mathcal{A}_{e}\cdot\mathcal{R}_{shareholder}\cdot\mathcal{A}_{p}$&0.08352&*\\
                    6&$\mathcal{A}_{e}\cdot\mathcal{R}_{subsidiary}\cdot\mathcal{A}_{e}\cdot\mathcal{R}_{control}\cdot\mathcal{A}_{p}$&0.08497&*\\
                    7&$\mathcal{A}_{e}\cdot\mathcal{R}_{subsidiary}\cdot\mathcal{A}_{e}\cdot\mathcal{R}_{produce}\cdot\mathcal{A}_{c}$&0.08632&*\\
                    8&$\mathcal{A}_{e}\cdot\mathcal{R}_{sales}\cdot\mathcal{A}_{e}\cdot\mathcal{R}_{produce}\cdot\mathcal{A}_{c}$&0.08756&*\\
                    9&$\mathcal{A}_{e}\cdot\mathcal{R}_{sales}\cdot\mathcal{A}_{e}\cdot\mathcal{R}_{control}\cdot\mathcal{A}_{p}$&0.09367&*\\
                   10&$\mathcal{A}_{e}\cdot\mathcal{R}_{shareholder}\cdot\mathcal{A}_{p}\cdot\mathcal{R}_{shareholder}\cdot\mathcal{A}_{e}$&0.09526&*\\
                   11&$\mathcal{A}_{e}\cdot\mathcal{R}_{subsidiary}\cdot\mathcal{A}_{e}\cdot\mathcal{R}_{manager}\cdot\mathcal{A}_{p}$ & 0.09831&*\\
                   12&$\mathcal{A}_{e}\cdot\mathcal{R}_{shareholder}\cdot\mathcal{A}_{p}\cdot\mathcal{R}_{control}\cdot\mathcal{A}_{p}$  &0.09836&*\\
                   13&$\mathcal{A}_{e}\cdot\mathcal{R}_{manager}\cdot\mathcal{A}_{p}\cdot\mathcal{R}_{manager}\cdot\mathcal{A}_{e} $&5.5101&-\\
                   14&$\mathcal{A}_{e}\cdot\mathcal{R}_{employee}\cdot\mathcal{A}_{p}\cdot\mathcal{R}_{manager}\cdot\mathcal{A}_{e}$&6.5260&-\\
                   15&$\mathcal{A}_{e}\cdot\mathcal{R}_{supply}\cdot\mathcal{A}_{e}\cdot\mathcal{R}_{control}\cdot\mathcal{A}_{p}$&9.7263&-\\
                   16&$\mathcal{A}_{e}\cdot\mathcal{R}_{shareholder}\cdot\mathcal{A}_{p}\cdot\mathcal{R}_{employee}\cdot\mathcal{A}_{p}$&14.4973&-\\
                   17&$\mathcal{A}_{e}\cdot\mathcal{R}_{sales}\cdot\mathcal{A}_{e}\cdot\mathcal{R}_{shareholder}\cdot\mathcal{A}_{p}$&23.5246&-\\
                   18&$\mathcal{A}_{e}\cdot\mathcal{R}_{supply}\cdot\mathcal{A}_{e}\cdot\mathcal{R}_{shareholder}\cdot\mathcal{A}_{p}$&27.7731&-\\
                   19&$\mathcal{A}_{e}\cdot\mathcal{R}_{control}\cdot\mathcal{A}_{p}\cdot\mathcal{R}_{employee}\cdot\mathcal{A}_{p}$&28.3672&-\\
                   20&$\mathcal{A}_{e}\cdot\mathcal{R}_{supply}\cdot\mathcal{A}_{e}\cdot\mathcal{R}_{manager}\cdot\mathcal{A}_{p}$&31.5267&-\\
\noalign{\smallskip}\hline
\end{tabular}
\begin{tablenotes}
 \footnotesize
 \item[1] *: p\textless 0.1, **: p\textless 0.05, ***: p\textless0.01, ****: p\textless 0.001
\end{tablenotes}
\end{threeparttable}
\end{table}

\subsection{Overall Comparisons of MP feature}\label{subsubsec:Accuracy of evaluation measures}
\begin{figure*}[h]
            \begin{subfigure}[t]{0.325\textwidth}
             \centering
            \includegraphics[width=\textwidth]{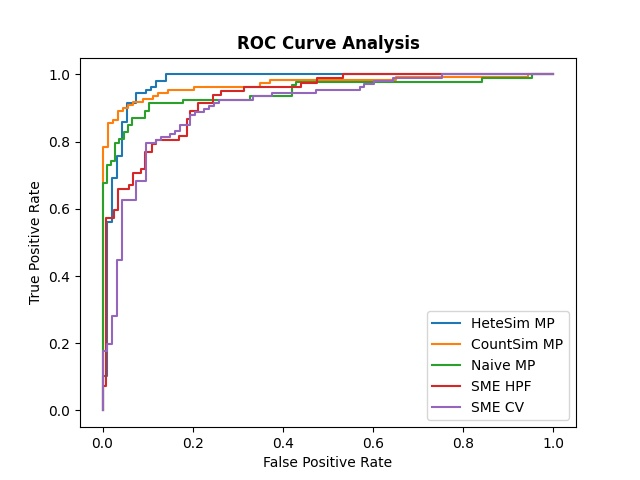} 
             \caption{the GEM dataset}
             \label{fig:auc1} 
             \end{subfigure}
              \begin{subfigure}[t]{0.325\textwidth}
               \centering
              \includegraphics[width=\textwidth]{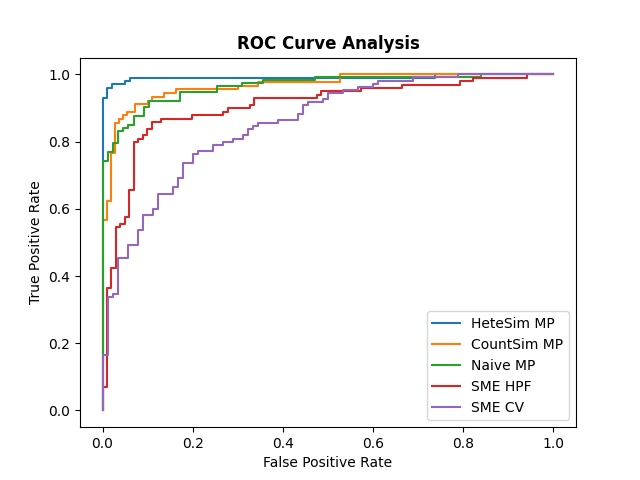}
               \caption{the STAR dataset}
             \label{fig:auc2} 
             \end{subfigure}
             \begin{subfigure}[t]{0.325\textwidth}
              \centering
              \includegraphics[width=\textwidth]{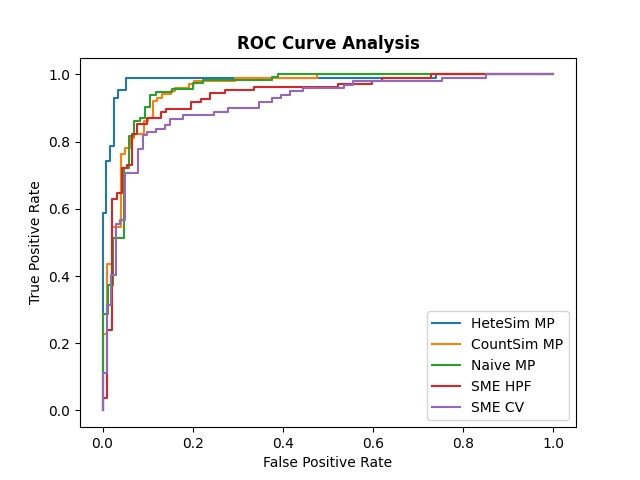}
               \caption{the SB dataset}
             \label{fig:auc3} 
              \end{subfigure}
                \caption{ROC curve for three datasets} 
\end{figure*}

In this section, we compare our three kinds of MP features with two kinds of other features proposed for evaluating SME credit risk.
One kind of the compared features is conventional features \cite{jrfm12010030},  such as current liquidity, quick ratio, assets turnover, a total of 16 financial indicators, and age of the enterprise, employment, a total of 5 non-financial indicators.
In our experiments, we call it \emph{SME CV}.
The other kind of the compared features is homogeneous path feature \cite{2017Bankruptcy} which is modeled from homogeneous information networks.
It contain only one object type and only one relation type, for example, two SMEs are related if they share a high-level manager.
In our experiments, we call it \emph{SME HPF}.
For our MP features, we respectively select the Naive MP features, CountSim MP features and HeteSim MP features according to the ranking result in Section \ref{subsec: mp feature selection} as the candidate features for comparison.
All the comparisons are still conducted on the mentioned three datasets.
To compare mentioned methods, we first select the top-10 performed features of each method.
And then we use their average AUC score as the overall score of each mentioned method.
The comparison results are summarized in Figures \ref{fig:auc1} - \ref{fig:auc3} and Table \ref{tab:auc}.

We can see that the heterogeneous MP features beats the SME CV features and the SME HPF features in all three datasets.
For the proposed MP features,  it turns out that: (1) All the MP features show better classification performance than the SME conventional features and the homogeneous path features; (2) The classification performance of the CountSim MP features and the HeteSim MP features beats the Naive MP features; (3)The classification performance of the CountSim MP features and the HeteSim MP features are similar. 
The above results demonstrate the effectiveness of our proposed features in classifying default SMEs.


\begin{table}
\caption{average AUC score comparison for three datasets}
\label{tab:auc}       
\centering
\begin{tabular}{cccccc}
  \hline\noalign{\smallskip}
       & SME CV&SME HPF &Naive MP&CountSim MP& HeteSim MP\\
     \noalign{\smallskip}\hline\noalign{\smallskip}      
       GEM& 0.716  & 0.728  &0.747&0.771&0.774\\
       STAR& 0.654  & 0.707 & 0.759& 0.767&0.791\\
       SB& 0.721 & 0.733 & 0.752&0.756&0.783 \\
  \noalign{\smallskip}\hline
\end{tabular}
\end{table}

\subsection{Discussion}\label{discussion}
\begin{figure*}[h]
            \begin{subfigure}[t]{0.325\textwidth}
             \centering
            \includegraphics[width=\textwidth]{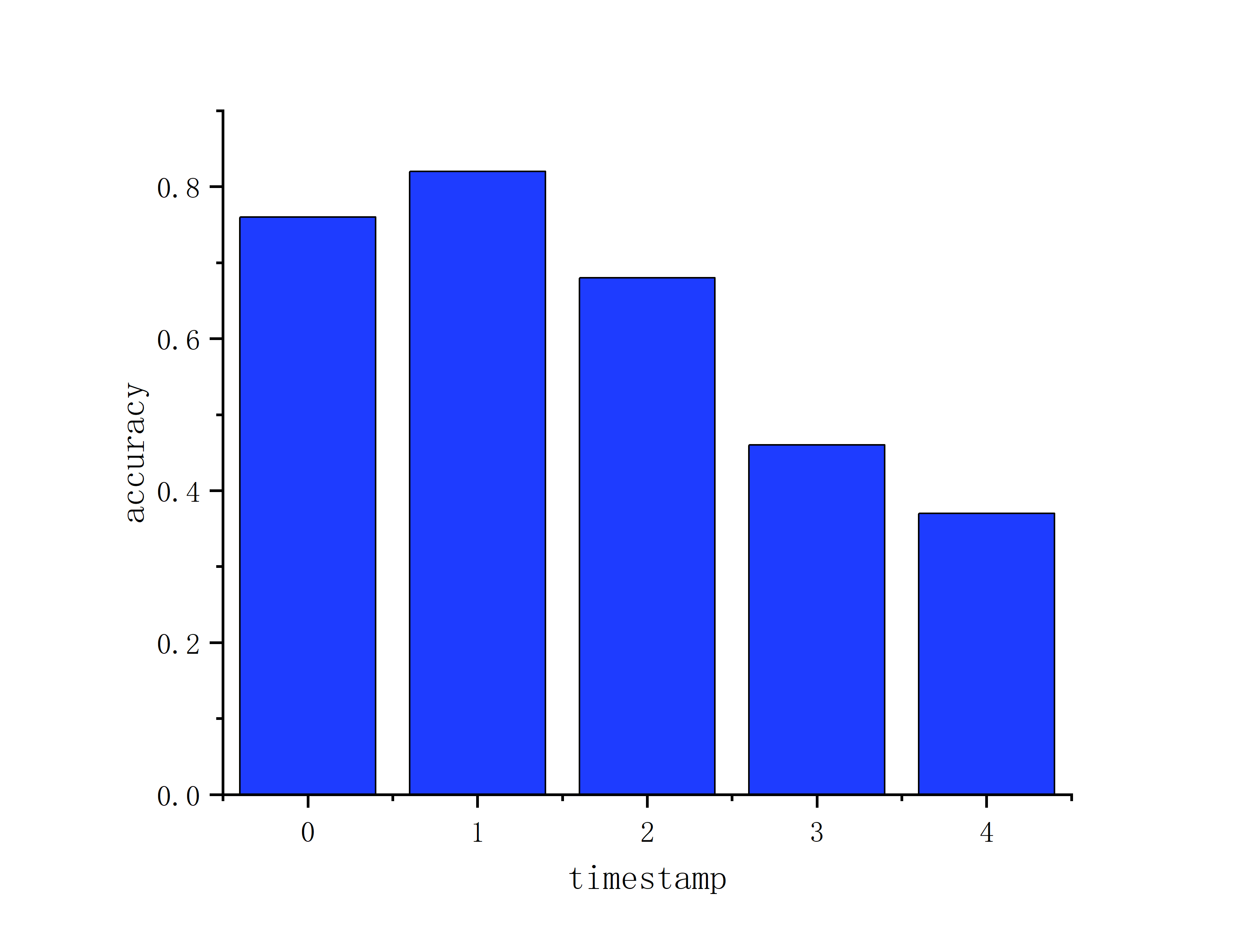} 
             \caption{the GEM dataset}
             \label{fig:time1} 
             \end{subfigure}
              \begin{subfigure}[t]{0.325\textwidth}
               \centering
              \includegraphics[width=\textwidth]{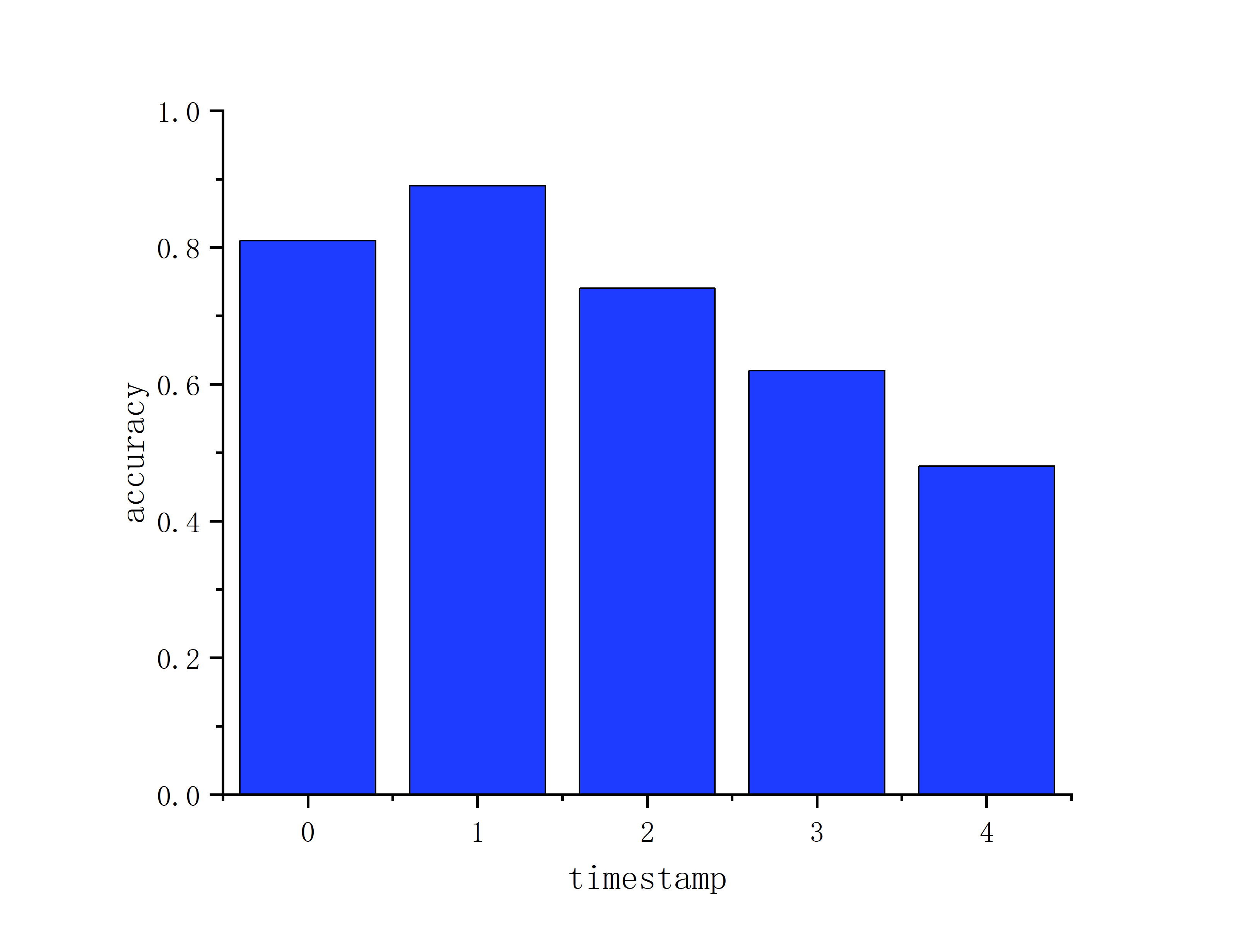}
               \caption{the STAR dataset}
             \label{fig:time2} 
             \end{subfigure}
             \begin{subfigure}[t]{0.325\textwidth}
              \centering
              \includegraphics[width=\textwidth]{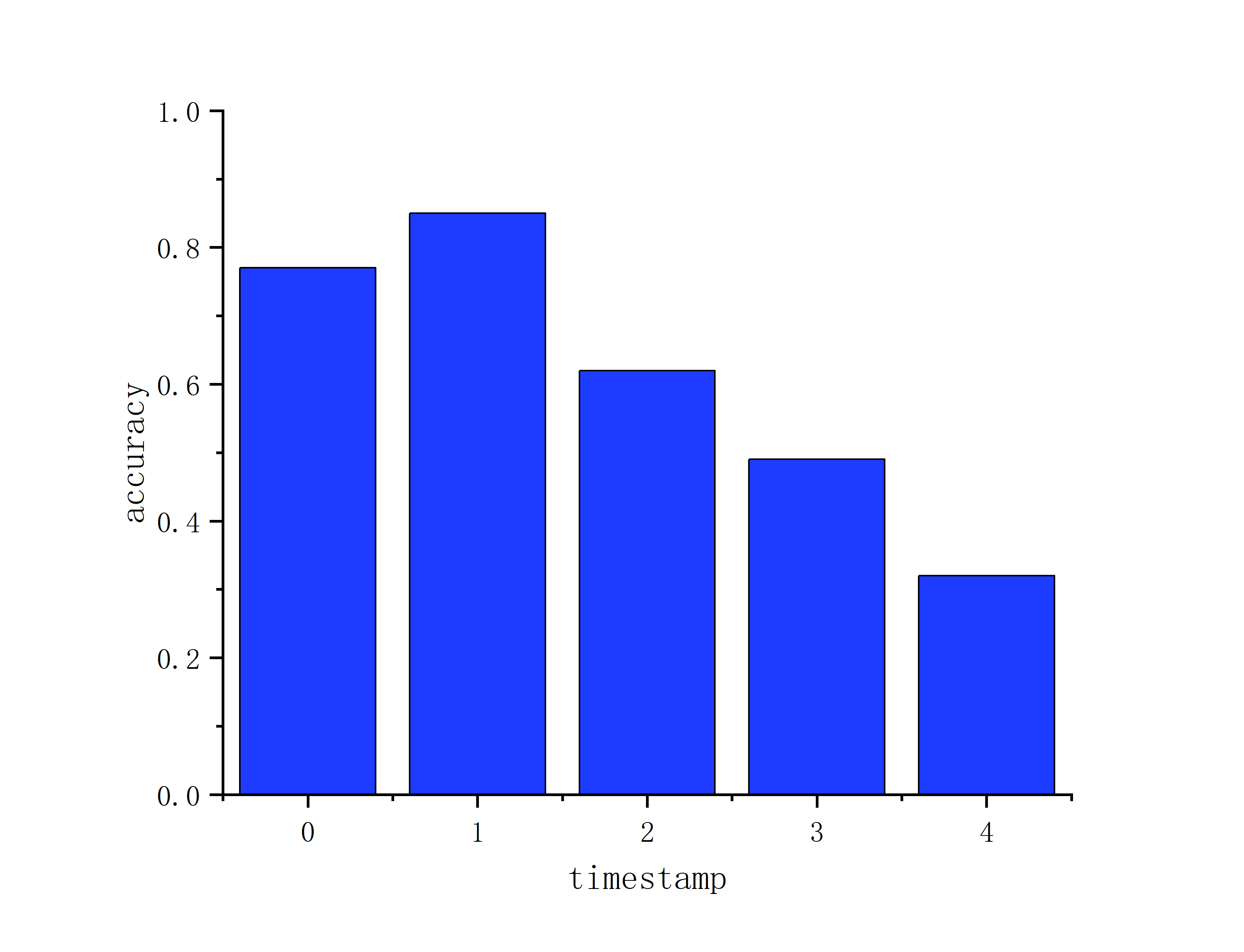}
               \caption{the SB dataset}
             \label{fig:time3} 
              \end{subfigure}
                \caption{ Classification accuracy of MP features under different timestamps} 
\end{figure*}

In this section, we will discuss some interesting point which we found in our experiments.
In general, prediction accuracy increases with data size increasing.
However, we found for SMEs the impact of data size is affected by the timestamp of data.
Next, we will detail and discuss how this affection comes.
Figure \ref{fig:time1} - \ref{fig:time3} shows the classification accuracy of meta path features under different timestamps.

It is interesting that when we extend SME data used in our model with the latest data in one year, the accuracy of the model increases.
But if we extend that with data before last year, the accuracy of the model shows a declining trend.
This phenomenon may be due to the fact that if the additional data is still in its valid duration, our model can be learnt more fully within the life circle of the enterprise.
But if the additional data is out of its valid duration, our model may be learnt out of the life circle and loss its effectiveness.
For example,  employee turnover rate over two years can not reflect the truth about the target enterprise now.

\section{Conclusion}\label{sec:Conclusion}
This paper proposes a meta path based SME credit risk evaluation method that models SME-related information as a heterogeneous information network.
In detail, we first build an SME heterogeneous information network based on four entity types and ten relation types.
The heterogeneous information network of SMEs can capture the relationship among related enterprises and provide more comprehensive and reliable information for the credit risk measurement of SMEs.
Then, we extracted meta path features associated with SME based on the information network schema, which represents the situation of SME credit risk.
Finally, we developed three features to evaluate the effect of meta path on SME credit risks.
The experimental result shows that our proposed SME credit risk measuring method has a higher significance than the conventional features and the homogeneous features.

\begin{acknowledgements}
This work is supported by the Project of Science and Technology Research and Development of China State Railway Group Co., Ltd. under Grant K2020Z002.
\end{acknowledgements}

%

\bibliographystyle{spmpsci}      
\bibliography{ref}   

\begin{thebibliography}{10}
\providecommand{\url}[1]{{#1}}
\providecommand{\urlprefix}{URL }
\expandafter\ifx\csname urlstyle\endcsname\relax
  \providecommand{\doi}[1]{DOI~\discretionary{}{}{}#1}\else
  \providecommand{\doi}{DOI~\discretionary{}{}{}\begingroup
  \urlstyle{rm}\Url}\fi

\bibitem{bauer2014hazard}
Bauer, J., Agarwal, V.: Are hazard models superior to traditional bankruptcy
  prediction approaches? a comprehensive test.
\newblock Journal of Banking \& Finance \textbf{40}, 432--442 (2014)

\bibitem{cultrera2016bankruptcy}
Cultrera, L., Br{\'e}dart, X.: Bankruptcy prediction: the case of belgian smes.
\newblock Review of Accounting and Finance  (2016)

\bibitem{edmister1972empirical}
Edmister, R.O.: An empirical test of financial ratio analysis for small
  business failure prediction.
\newblock Journal of Financial and Quantitative analysis \textbf{7}(2),
  1477--1493 (1972)

\bibitem{2020Bankruptcy}
Gang, K.A., Yong, X.A., Yi, P.B., Feng, S.C., Yang, C.A., Kc, D., Sk, D.:
  Bankruptcy prediction for smes using transactional data and two-stage
  multiobjective feature selection.
\newblock Decision Support Systems \textbf{140} (2020)

\bibitem{gupta2017heteclass}
Gupta, M., Kumar, P., Bhasker, B.: Heteclass: A meta-path based framework for
  transductive classification of objects in heterogeneous information networks.
\newblock Expert Systems with Applications \textbf{68}, 106--122 (2017)

\bibitem{2013Feature}
Hajek, P., Michalak, K.: Feature selection in corporate credit rating
  prediction.
\newblock Knowledge-Based Systems \textbf{51}, 72--84 (2013)

\bibitem{hosseini2018heteromed}
Hosseini, A., Chen, T., Wu, W., Sun, Y., Sarrafzadeh, M.: Heteromed:
  Heterogeneous information network for medical diagnosis.
\newblock In: Proceedings of the 27th ACM International Conference on
  Information and Knowledge Management, pp. 763--772 (2018)

\bibitem{2019Cash}
Hu, B., Zhang, Z., Shi, C., Zhou, J., Qi, Y.: Cash-out user detection based on
  attributed heterogeneous information network with a hierarchical attention
  mechanism.
\newblock Proceedings of the AAAI Conference on Artificial Intelligence
  \textbf{33} (2019)

\bibitem{jamali2010matrix}
Jamali, M., Ester, M.: A matrix factorization technique with trust propagation
  for recommendation in social networks.
\newblock In: Proceedings of the fourth ACM conference on Recommender systems,
  pp. 135--142 (2010)

\bibitem{ji2010graph}
Ji, M., Sun, Y., Danilevsky, M., Han, J., Gao, J.: Graph regularized
  transductive classification on heterogeneous information networks.
\newblock In: Joint European Conference on Machine Learning and Knowledge
  Discovery in Databases, pp. 570--586. Springer (2010)

\bibitem{2010Predicting}
Lugovskaya, L.: Predicting default of russian smes on the basis of financial
  and non-financial variables.
\newblock Journal of Financial Services Marketing \textbf{14}(4), 301--313
  (2010)

\bibitem{ma2009learning}
Ma, H., King, I., Lyu, M.R.: Learning to recommend with social trust ensemble.
\newblock In: Proceedings of the 32nd international ACM SIGIR conference on
  Research and development in information retrieval, pp. 203--210 (2009)

\bibitem{ma2008sorec}
Ma, H., Yang, H., Lyu, M.R., King, I.: Sorec: social recommendation using
  probabilistic matrix factorization.
\newblock In: Proceedings of the 17th ACM conference on Information and
  knowledge management, pp. 931--940 (2008)

\bibitem{moro2013loan}
Moro, A., Fink, M.: Loan managers’ trust and credit access for smes.
\newblock Journal of banking \& finance \textbf{37}(3), 927--936 (2013)

\bibitem{popescul2003statistical}
Popescul, A., Ungar, L.H.: Statistical relational learning for link prediction.
\newblock In: IJCAI workshop on learning statistical models from relational
  data, vol. 2003. Citeseer (2003)

\bibitem{2010Evaluation}
Psillaki, M., Tsolas, I.E., Margaritis, D.: Evaluation of credit risk based on
  firm performance.
\newblock European Journal of Operational Research \textbf{201}(3), 873--881
  (2010)

\bibitem{jrfm12010030}
Ptak-Chmielewska, A.: Predicting micro-enterprise failures using data mining
  techniques.
\newblock Journal of Risk and Financial Management \textbf{12}(1) (2019).
\newblock \doi{10.3390/jrfm12010030}.
\newblock \urlprefix\url{https://www.mdpi.com/1911-8074/12/1/30}

\bibitem{sermpinis2018modelling}
Sermpinis, G., Tsoukas, S., Zhang, P.: Modelling market implied ratings using
  lasso variable selection techniques.
\newblock Journal of Empirical Finance \textbf{48}, 19--35 (2018)

\bibitem{shi2014hetesim}
Shi, C., Kong, X., Huang, Y., Philip, S.Y., Wu, B.: Hetesim: A general
  framework for relevance measure in heterogeneous networks.
\newblock IEEE Transactions on Knowledge and Data Engineering \textbf{26}(10),
  2479--2492 (2014)

\bibitem{2016A}
Shi, C., Li, Y., Zhang, J., Sun, Y., Yu, P.S.: A survey of heterogeneous
  information network analysis.
\newblock IEEE Transactions on Knowledge and Data Engineering \textbf{29}(1),
  17--37 (2016)

\bibitem{shi2015semantic}
Shi, C., Zhang, Z., Luo, P., Yu, P.S., Yue, Y., Wu, B.: Semantic path based
  personalized recommendation on weighted heterogeneous information networks.
\newblock In: Proceedings of the 24th ACM International on Conference on
  Information and Knowledge Management, pp. 453--462 (2015)

\bibitem{sun2012relation}
Sun, Y., Aggarwal, C.C., Han, J.: Relation strength-aware clustering of
  heterogeneous information networks with incomplete attributes.
\newblock arXiv preprint arXiv:1201.6563  (2012)

\bibitem{sun2013mining}
Sun, Y., Han, J.: Mining heterogeneous information networks: a structural
  analysis approach.
\newblock Acm Sigkdd Explorations Newsletter \textbf{14}(2), 20--28 (2013)

\bibitem{sun2012will}
Sun, Y., Han, J., Aggarwal, C.C., Chawla, N.V.: When will it happen?
  relationship prediction in heterogeneous information networks.
\newblock In: Proceedings of the fifth ACM international conference on Web
  search and data mining, pp. 663--672 (2012)

\bibitem{sun2009ranking}
Sun, Y., Yu, Y., Han, J.: Ranking-based clustering of heterogeneous information
  networks with star network schema.
\newblock In: Proceedings of the 15th ACM SIGKDD international conference on
  Knowledge discovery and data mining, pp. 797--806 (2009)

\bibitem{tian2015variable}
Tian, S., Yu, Y., Guo, H.: Variable selection and corporate bankruptcy
  forecasts.
\newblock Journal of Banking \& Finance \textbf{52}, 89--100 (2015)

\bibitem{2017Bankruptcy}
Tobback, E., Bellotti, T., Moeyersoms, J., Stankova, M., Martens, D.:
  Bankruptcy prediction for smes using relational data.
\newblock Decision Support Systems \textbf{102}(oct.), 69--81 (2017)

\bibitem{tsai2017risk}
Tsai, M.F., Wang, C.J.: On the risk prediction and analysis of soft information
  in finance reports.
\newblock European Journal of Operational Research \textbf{257}(1), 243--250
  (2017)

\bibitem{wang2016text}
Wang, C., Song, Y., Li, H., Zhang, M., Han, J.: Text classification with
  heterogeneous information network kernels.
\newblock In: Thirtieth AAAI Conference on Artificial Intelligence (2016)

\bibitem{wang2018shine}
Wang, H., Zhang, F., Hou, M., Xie, X., Guo, M., Liu, Q.: Shine: Signed
  heterogeneous information network embedding for sentiment link prediction.
\newblock In: Proceedings of the Eleventh ACM International Conference on Web
  Search and Data Mining, pp. 592--600 (2018)

\bibitem{2013Recommendation}
Xiao, Y., Xiang, R., Sun, Y., Sturt, B., Han, J.: Recommendation in
  heterogeneous information networks with implicit user feedback.
\newblock In: Acm Conference on Recommender Systems (2013)

\bibitem{yin2020evaluating}
Yin, C., Jiang, C., Jain, H.K., Wang, Z.: Evaluating the credit risk of smes
  using legal judgments.
\newblock Decision Support Systems \textbf{136}, 113364 (2020)

\bibitem{2013Modeling}
Zhong, E., Wei, F., Yin, Z., Qiang, Y.: Modeling the dynamics of composite
  social networks.
\newblock In: Proceedings of the 19th ACM SIGKDD international conference on
  Knowledge discovery and data mining (2013)

\end{thebibliography}

%
%

\end{document}